\documentclass[referee]{aa} 
\usepackage{graphicx}
\usepackage{txfonts}

\begin{document}
   \title{Flaming, bright galaxies along the filaments of A\,2744}

   \subtitle{}

   \author{F. Braglia
          \inst{1}
          \and
          D. Pierini\inst{1}
	  \and
	  H. B\"{o}hringer\inst{1}
          }

   \offprints{F. Braglia}

   \institute{Max-Planck-Institut f\"{u}r extraterrestrische Physik (MPE), 
              Giessenbachstrasse 1, D-85748 Garching\\
              \email{fbraglia@mpe.mpg.de; dpierini@mpe.mpg.de; hxb@mpe.mpg.de}
             }


 
  \abstract
   {The existence of a clumpy, large-scale filamentary structure is at the basis
of the current paradigm of cosmic formation and evolution of clusters.
The star-formation history of galaxies falling into a cluster is altered
as a result of the environmental stresses.}
   {We investigate the relation between substructure and properties
of the galaxy population in a $30^{\prime} \times 30^{\prime}$ region
centered on the massive merging cluster A\,2744 at $z \sim 0.3$.}
   {Multi-object spectroscopy at low resolution and BVR photometry
are used to detect the presence of substructure through a Dressler--Schectman
analysis and the photometric redshift technique, respectively. Galaxies at the same
photometric redshift of the cluster are divided into red and blue
according to their distribution in the B-R vs. R colour--magnitude diagram.}
   {We identify two large-scale filaments associated with A\,2744.
Along these filaments, the blue-to-red galaxy number ratio increases
together with the cluster-centric distance but peaks slightly beyond
the cluster virial radius. The filaments host a population of bright, large
(i.e. more luminous than $\rm R^{\star}$ for the main body of the cluster
and with angular sizes of 13--22 $h_{70}^{-1}$ kpc) blue galaxies
that is hardly found among galaxies lying in a neighbouring low-density environment
at the same redshift of the cluster.}
   {These results can be interpreted as a manifestation of galaxy harassment.}


   \maketitle
%

\section{Introduction}

Detailed numerical simulations show that clusters form at the intersection
of filaments and sheets of matter in the evolving large-scale structure
of the Universe (e.g., Colberg et al. \cite{colberg99}).
These infall-pattern features are correlated in time and appear to be clumpy
rather than homogeneous.
Thus they define preferred directions from which clusters are feeded
with lumps of matter.

The existence of extended filamentary structures and voids in the large-scale
spatial distribution of local galaxies is known since the advent
of redshift surveys (Davis et al. \cite{davis82}).
Later X-ray observations detected the emission from the hot (10$^5$--10$^7$ K)
gas in filaments of nearby clusters (Briel \& Henry \cite{briel95};
Kull \& B\"{o}hringer \cite{kull99}; Scharf et al. \cite{scharf00};
Durret et al. \cite{durret03}).

In addition, optical observations established the existence
of a correlation between the morphology or star-formation rate
of a galaxy and the local galaxy density or cluster-centric distance
(Dressler \cite{dressler80}; Whitmore et al. \cite{whitmore93};
Dressler et al. \cite{dressler97}; Balogh et al. \cite{balogh99};
Poggianti et al. \cite{poggianti99}).
In particular, the spectral-index analysis for galaxies of the CNOC1 cluster
sample at $z \sim 0.3$ demonstrated that the radial increase
in star-formation activity means that the last episode of star formation
occurred more recently in galaxies farthest from the cluster center
(Balogh et al. \cite{balogh99}).
Different mechanisms can establish this pattern and, thus,
the observed difference in star-formation activity
between cluster and field galaxies (e.g., Abraham et al. \cite{abraham96};
Morris et al. \cite{morris98}; Balogh et al. \cite{balogh99}).
They include ram-pressure stripping by the intracluster medium (ICM)
(e.g., Gunn \& Gott \cite{gunn72}; Quilis et al. \cite{quilis00}),
close encounters (Barnes \cite{barnes92}), ``galaxy harassment''
(Moore et al. \cite{moore96}) and ``strangulation''
(Larson et al. \cite{larson80}).

Determining the relative importance of these mechanisms requires
probing galaxy properties as a function of the infall-region topology
and mass of a cluster.
In this pilot study, we investigate A\,2744
(AC\,118 or RXCJ0014.3-3022) out of the REFLEX-DXL
(Distant X-ray Luminous) catalogue.
This is a homogeneous, unbiased sample drawn from the REFLEX survey
(B\"{o}hringer et al. \cite{boehringer01}) comprising the thirteen
most luminous clusters at $z = 0.27$--0.31 in the Southern emisphere,
with X-ray ([0.1-2.4] keV) luminosities in excess
of $\rm 10^{45}~h_{70}^{-1}~erg~s^{-1}$ (Zhang et al. \cite{zhang06}).
Multiple evidence shows that the main body of A\,2744
is a merging system (Boschin et al. \cite{boschin06} and references therein).
In a preliminary analysis (B\"ohringer at al. \cite{boehringer06}),
the comparison of X-ray and optical properties suggested the presence
of two filamentary structures stemming out of the main body of the cluster.
Here we present a more accurate kinematical analysis of the cluster
and investigate the galaxy populations associated with these filaments.

%
\section{Data description and reduction}

Imaging in the B-, V-, and R-passbands was carried out using the wide-field
imager (WFI) on the ESO/MPG-2.2m telescope at La Silla, Chile
in September 2000, with a seeing of $\sim 1^{\prime\prime}$.
The WFI data were reduced using the data reduction system developed
for the ESO Imaging Survey (EIS, Renzini \& da Costa \cite{renzini97})
and its associated EIS/MVM image processing library version 1.0.1
({\it Alambic}, Vandame \cite{vandame04})\footnote{{\it Alambic} is
a publicly available software designed to automatically transform raw images
from single/multi-chip optical/infrared cameras into reduced images
for scientific use.}.
Source detection and photometry were performed with {\em SExtractor}
(Bertin \& Arnouts \cite{bertin96}).
Source photometry was extracted in fixed circular apertures
($2^{\prime\prime}$ wide for the determination of photometric redshifts)
or in a flexible (Kron-like, Kron \cite{kron80}) elliptical aperture
with a Kron-factor of 2.5.
Magnitudes were calibrated to the Johnson--Cousins system using standard stars
(Landolt \cite{landolt92}), corrected for galactic extinction
(Schlegel et al. \cite{schlegel98}), and expressed in the AB system.
The photometric catalogue is complete down to $\rm R \sim 23.5$
and contains objects across a field of $1/4$ of a square degree.

Multi-object spectroscopy was performed with VLT-VIMOS in low resolution
($R = 200$) mode (LR-Blue grism) in September 2004.
With a slit-width of $1^{\prime\prime}$, the expected uncertainty on the observed
velocities is 250--300 km s$^{-1}$. Comparison with previously known redshifts 
in the same region gives a mean error of 276 km s$^{-1}$, 
in agreement with with the value estimated by Le F\`{evre} et al. \cite{lefevre05} for the same LR-Blue grism.
This is enough to establish the membership of a galaxy
and the presence of large-scale structure for a massive cluster
like A\,2744.
Objects with $\rm I \le 22.5$ were selected as targets for spectroscopy
from VLT-VIMOS pre-imaging of a $24^{\prime} \times 22^{\prime}$ field.
This limiting magnitude corresponds approximatively to an $\rm I^{\star}+3$
galaxy at the redshift of the cluster (0.3068, see Couch et al. \cite{couch98}).
The adopted selection criterion prevents a bias against cluster members
with specific star-formation histories (i.e., colours).
Thus, it also prevents the completeness of the spectroscopic sample
from being dependent on the local galaxy density, given the known relation
between this density and the star-formation rate of a galaxy.
The spectroscopic observations provided about 900 spectra.
These data were reduced using the dedicated software VIPGI\footnote{VIPGI
(VIMOS Interactive Pipeline and Graphical Interface) (Scodeggio et al.
\cite{scodeggio05}) is developed by the VIRMOS Consortium to handle
the reduction of the VIMOS data for the VVDS (VIMOS VLT Deep Survey, Le F\`{e}vre et al. \cite{lefevre05}).}.

Throughout the paper we adopt a $\rm \Lambda$CDM cosmology where $\Omega_{\rm m} = 0.3$,
$\Omega_{\rm \Lambda} = 0.7$, and $h_{70} = H_0/~{\rm 70~km~s^{-1}~Mpc^{-1}} = 1$.

%
\section{Analysis}

\subsection{Evidence of large-scale structure}

Spectroscopic redshifts were first converted into velocities according to
Danese et al. (\cite{danese80}).
We then applied a recursive 3$\sigma$ clipping algorithm to the subsample
of galaxies selected in a velocity interval of 10,000 km s$^{-1}$ width
centered on the known redshift of the cluster.
As a result, we identified 134 cluster members.
This number was increased by additional 60 cluster members
identified from redshifts available in the {\it NED}, mostly lying
within 1 Mpc from the center of the cluster.
The ensuing mean redshift of the cluster (0.3068$\pm$0.006)
is in excellent agreement with previous determinations.
The total number of 194 spectroscopic cluster members is twice
as large as the number considered in the analysis
of Boschin et al. (\cite{boschin06}).
Furthermore, it probes an area almost four times as large.

We determine a rest-frame velocity dispersion of the cluster
equal to 1509 km s$^{-1}$.
If A\,2744 were a dynamically relaxed system, its velocity
dispersion and virial radius\footnote{$R_{\rm Vir} = 2.5~h_{70}^{-1}$ Mpc,
from the X-ray analysis of Zhang et al. (\cite{zhang06}).} would give
a dynamical mass equal to $3.3 \times 10^{15}~h_{70}^{-1}~\rm M_{\odot}$.
However, a Kolmogorov--Smirnov (KS) test reveals that the distribution
of radial velocities has only a 0.02\% probability to be gaussian.
This is consistent with A\,2744 being a merging system.
Combining galaxy velocity and position information (Dressler \& Schectman
\cite{dressler88}), we identify two very prominent substructures,
lying at 9.3$^\prime$ to the NW and 8.2$^\prime$ to the S with respect to the cluster main body.
These substructures are external but contiguous to those identified
by Boschin et al. (\cite{boschin06}).
They find counterparts in the structures seen in the hot-gas entropy map
(Finoguenov et al. \cite{finoguenov05}) as remarked
in B\"{o}hringer et al. (\cite{boehringer06}).
In particular (see Fig. 1), we find that the NW structure corresponds to
a clump of 23 objects with a velocity distribution characterized by
a mean relative velocity of +566 km s$^{-1}$, a dispersion of 428 km s$^{-1}$
and a positive skew.
We picture these galaxies as infalling from the near side of the cluster.
Conversely, the S structure corresponds to a tighter clump of 15 objects,
with a mean relative velocity of -738 km s$^{-1}$ and a dispersion
of 318 km s$^{-1}$.
This suggests infall from the far side of the cluster.

We run extended Monte Carlo simulations over our spectroscopic sample, reshuffling the redshifts of the objects while mantaining their positions as in Dressler \& Schectman (\cite{dressler88}) and adding a random gaussian error with sigma equal to the dispersion found. We also run the same test by generating a gaussian velocity distribution.
Comparison of the DS-test deviation distributions shows that the inclusion of errors is not sufficient to change the overall behavior: the two distributions are always consistent, the mean KS statistical parameter being 0.57, to be confronted with a rejection threshold of 0.88. Conversely, comparison of the true redshift distribution with the gaussian case shows that the two distributions differ not only in velocity, but also in the distribution of DS deviations (we obtain a KS parameter of 0.81, against a threshold for rejection of 0.53).
The spatial location of the highly deviating peaks does not change position, so they are clearly robust against the size of velocity errors. This is mainly due to the statistical nature of the DS test, which looks for deviations of whole clumps of galaxies by comparing the velocity mean and dispersion of individual groups with the overall velocity dispersion of the cluster. So the effect of errors of individual velocities is smeared down.

   \begin{figure}
   \centering
   \includegraphics[width=9 cm]{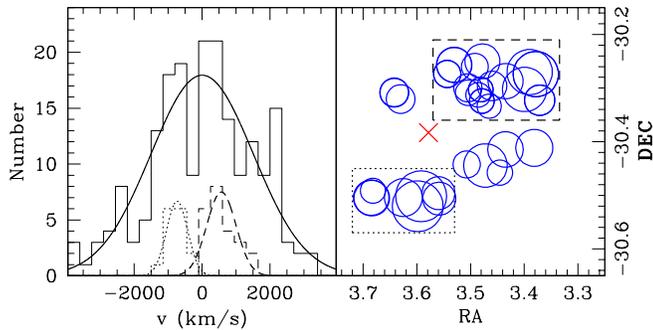}
      \caption{Left: velocity distribution for different regions of A\,2744, i.e.: the overall system ({\it solid line}), with a velocity dispersion of 1509 km s$^{-1}$; the substructure to the NW ({\it dashed line}), with a mean velocity of 566 km s$^{-1}$ and a dispersion of 428 km s$^{-1}$; the substructure to the S ({\it dotted line}), with a mean velocity of -738 km s$^{-1}$ and a dispersion of 314 km s$^{-1}$. Right: Dressler-Schectman test for the spectroscopic cluster members; the dashed- and dotted-line boxes outline respectively the NW and S substructures detected as very significant deviations from the overall velocity distribution. The red cross cross marks the cluster X-ray center.}
      \label{Fig.1}
   \end{figure}

\begin{figure}
   \centering
   \includegraphics[width=8 cm]{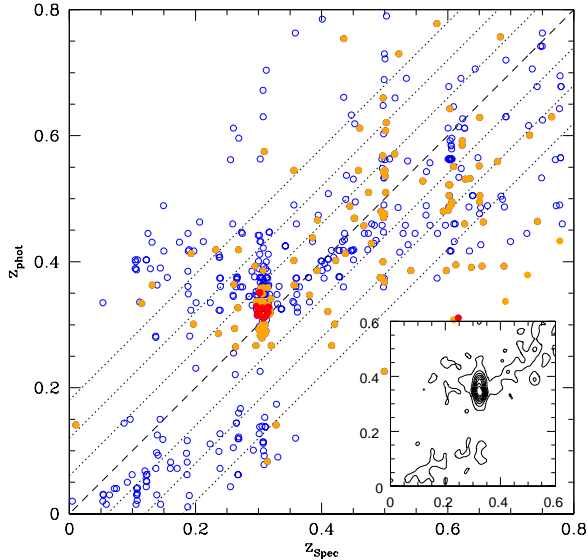}
   \caption{Comparison of spectroscopic and photometric redshifts. Orange dots: faint red galaxies; red dots: bright red galaxies; blue circles: blue galaxies. The dashed lines show the 1$\sigma$, 2$\sigma$ and 3$\sigma$ intervals, where we take as $\sigma$ the spread of the photo-$z$ at the cluster redshift. The lower right insert shows a region of the same distribution as a contour plot to highlight the density (almost 200 points) in the cluster region.}
   \label{Fig.2}
\end{figure}

In order to map these structures beyond the area covered by spectroscopy,
we determine photometric redshifts for the whole photometric catalogue
with the {\em HyperZ} code (Bolzonella et al. \cite{bolzonella00}).
Photo-$z$ solutions are trained on the available spectroscopic redshifts (about 800 non-stellar objects in the redshift range [0,0.85]), so to select the most efficient templates at the cluster redshift. We divide our spectro-photometric sample into red and blue objects on the basis of the cluster red sequence as described in Sect. 3.2. Then, we separate objects into bright and faint, where we define as ``bright'' the galaxies that are equally or more luminous than the characteristic magnitude for the main body of the cluster, i.e., $\rm R^{\star} \sim 19.6 \pm 0.3$ (Busarello et al. \cite{busarello02}). For the resulting 4 catalogs we check independently which templates to use for the photo-$z$ determination, running {\em HyperZ} over the same redshift range. Galaxies at the distance of the cluster are then identified as objects in the photo-$z$ range [0.28,0.40]. This corresponds to the 1$\sigma$ interval centered on the mean photometric redshift (0.34) of the spectroscopic cluster members. The rms uncertainty is 0.06 at the spectroscopic redshift of A\,2744. We see a somewhat wider spread for blue objects, as can be expected; nevertheless, the cluster stands clearly out in the photo-$z$ distribution (see Fig.~\ref{Fig.2}). The $\chi^2$ distribution for random objects from the samples exhibits always a single minimum (and thus a single solution).

We check the occurrence of catastrophic failures (i.e., wrong identifications) running {\em HyperZ} over the spectro-photometric catalogue with increasing redshift ranges (from [0,0.85] to [0,6]). This is done for both red and blue objects independently. No catastrophic failure is found up to $z \sim 1.5$. From this result and the relatively bright magnitude limit of the sample ($\rm R \sim 23.5$), we adopt a safe range of [0,1] for the photo-$z$ determination across the whole field.

To check the robustness of this assumption against contamination from high-$z$ outliers, we apply the same photometric redshift analysis over a simulated catalogue of objects in the chosen magnitude range and with a suitable redshift distribution. A sample of $\sim$30000 mock objects is thus created with the {\em HyperZ} routine {\em make\_catalog} down to $\rm R = 23.5$ and with a redshift distribution consistent with deep, magnitude-selected spectroscopic surveys (VVDS-Deep, Le F\`{e}vre et al. \cite{lefevre05}) to replicate as much as possible a natural redshift distribution over the field of view.

\begin{figure}
   \centering
   \includegraphics[width=8 cm]{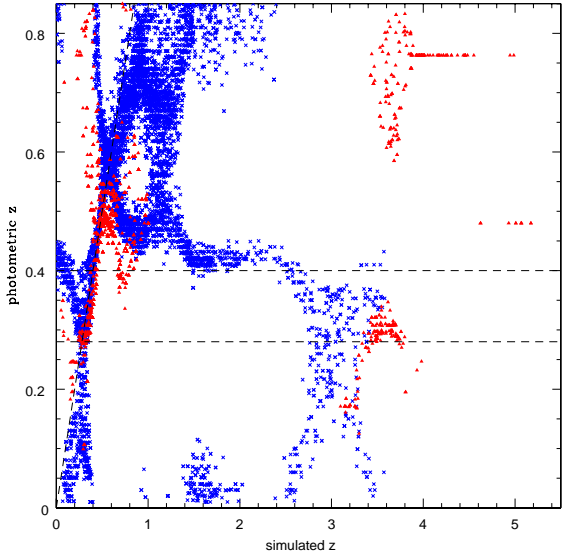}
   \caption{Simulated contamination from high-$z$ outliers. Red triangles represent old, passively evolving galaxies; blue crosses, star-forming galaxies. The dashed lines mark the cluster region in redshift space. Three main sources of contamination are identified: high-$z$ red galaxies (105 objects, contamination of 6.9\%); high-$z$ blue galaxies (185 objects, contamination of 1.7\%); and foreground blue galaxies (349 objects, contamination of 3.2\%).}
   \label{Fig.3}
\end{figure}

We run {\em HyperZ} with the selected templates over the simulated catalogue constrained to the redshift range [0,1]; the result is shown in Fig.~\ref{Fig.3}. While we see (as expected) some contamination from outliers, the wrong identifications mainly lie outside the photometric redshift range of the cluster (0.28--0.4).

In this simulation we did not take into account the presence of the cluster at $z = 0.3068$, as we want to assess the pure contamination from background objects. Comparing the number of interlopers with the photometric cluster members (5759 objects within the cluster photo-$z$ range) shows that the contamination is still quite low (1.8\% for high-$z$ red objects, 3.2\% for high-$z$ blue objects and 6.1\% for blue foreground objects). Thus, we conclude that the effect of contamination in the cluster redshift range is negligible.

\begin{figure}
   \centering
   \includegraphics[width=8 cm]{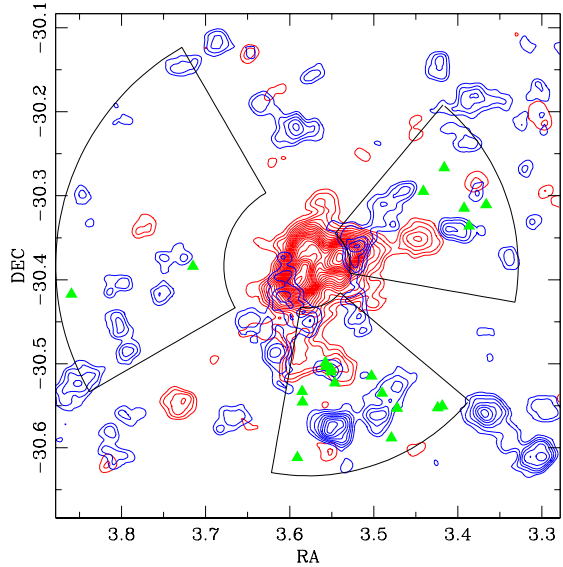}
   \caption{Density contours for red and blue galaxies at the distance (photo-$z$) of the cluster, in multiples of background RMS. The red objects density map shows the unrelaxed main body of the cluster and the head of the filamentary structures towards the S and NW. The blue objects map shows the presence of regions of high density in the outskirts of the cluster. In particular, two high density regions toward the S and NW lie on the extension of the substructures emerging from the cluster main body. The regions delimiting the two filaments and the ``field'' are also shown. Triangles mark the position of the luminous blue galaxies.}
              \label{Fig.4}
    \end{figure}

For the 5759 objects at the distance of the cluster, divided
into a red subsample and a blue one (see Sect. 3.2), we determine separate
number density maps.
As Fig.~\ref{Fig.4} shows, the bulk of the red objects sits in the main body
of the cluster, as expected.
However, two extensions of red objects are clearly seen towards
the same NW and S directions defined by the structures identified
by the Dressler--Schectman analysis.
On the other hand, the blue objects are almost evenly spread across the imaged region.
A closer look reveals a chain of highly significant overdensities
of blue objects stemming out of the S extension of red objects
and turning to the SW.
This is very suggestive of an extended, clumpy filament as simulations predict.
An extended filament in the NW direction is less evident.

\subsection{Star formation along the large-scale structure}

As a zero-point for the description of the star-formation activity
in the region of A\,2744, we take the red sequence
of the likely massive, old, passively evolving galaxies
in the B-R vs. R colour--magnitude diagram.
Here it is defined as the locus of galaxies with $\rm 17 \le R \le 20$
and B-R $\sim 2.2$ within a cluster-centric distance of 3$^\prime$
(i.e., about 800 kpc).
Then we divide the total sample of 5759 galaxies at the distance of the cluster
into a red subsample and a blue one.
An object is classified as blue if its B-R colour is bluer
than that of a mean red-sequence object with the same R-magnitude
at more than the 3$\sigma$ level.

Furthermore, we define two circular sectors with a 60$\rm ^o$ aperture
and radii from 3$^\prime$ to 15$^\prime$ which encompass the two filaments.
A third circular sector with a 90$\rm ^o$ aperture and radii from 6$^\prime$
to 18$^\prime$ defines an area equivalent to the sum of the previous
two regions where no structure is evident.
This third region represents the comparison ``field'' at the same redshift
of the cluster.
Down to the completeness limit, there are 570 and 500 objects
in the S and NW circular sectors, respectively,
against a total of 796 in the field.
Hence the significance of the average overdensity is equal to 9$\sigma$
and 5$\sigma$ for the S and NW circular sectors, respectively,
and to 10$\sigma$ overall.
The overdensity is driven by faint galaxies: 15$\sigma$ and 11$\sigma$
for the red and blue galaxies with $\rm R^{\star} < R \le 23.5$,
respectively, against 6$\sigma$ and 5$\sigma$ for the red and blue galaxies
with $\rm R \leq R^{\star}$, respectively.

\begin{figure}
   \centering
   \includegraphics[width=9 cm]{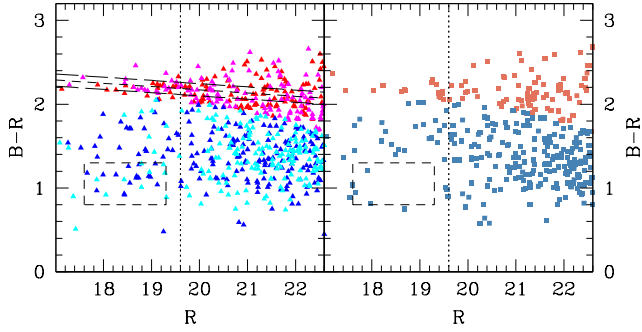}
      \caption{Colour-magnitude diagram for the galaxies in the filaments (left) or in the field (right). The population of the two filaments is represented in different colors: red objects appear in red (S filament) and magenta (NW filament), blue objects in blue (S filament) and cyan (NW filament). The signature of a red sequence can clearly be seen for the filaments, as well as the peculiar population of blue galaxies, in the dashed box; this population is not matched in the field. In each panel, the dashed vertical line marks the value of $\rm R^\star$ for the main body of the cluster.}
         \label{Fig.5}
   \end{figure}

Figure~\ref{Fig.5} reproduces the B-R vs. R colour--magnitude diagram
for galaxies in these three regions.
Luminous, red galaxies exhibit similar, tight colour-magnitude relations.
In the higher-density regions, red galaxies with $17 \le R \le 20$
define a red-sequence that is shifted bluewards by $0.012 \pm 0.003$ mag
with respect to the red-sequence in the main body of the cluster.
As for the population of blue galaxies, a KS-test run for bins of 0.5 R-mag
and the full B-R colour range shows a very significant (9$\sigma$)
overabundance of galaxies with $\rm 17.6 \leq R \leq 19.3$
and $\rm 0.8 \leq$ B-R $\leq 1.3$ in the higher-density regions.
This does not simply mirror the overabundance of luminous, blue galaxies there.

These bright, blue (and, thus, star-forming) galaxies are large systems,
as confirmed visually by eye-ball estimation of their angular sizes 
to about 3--$5^{\prime\prime}$ (i.e. about 13--22 $h_{70}^{-1}$ kpc).
They lie mostly in or close to overdensities of blue objects and around or beyond
the cluster virial radius $R_{\rm Vir}$ (about $9^{\prime}$ or 2.5 Mpc).
Out of these 19 luminous, blue galaxies, only three can be part
of interacting pairs, whereas all are surrounded by several small companions.

   \begin{figure}
   \centering
   \includegraphics[width=9 cm]{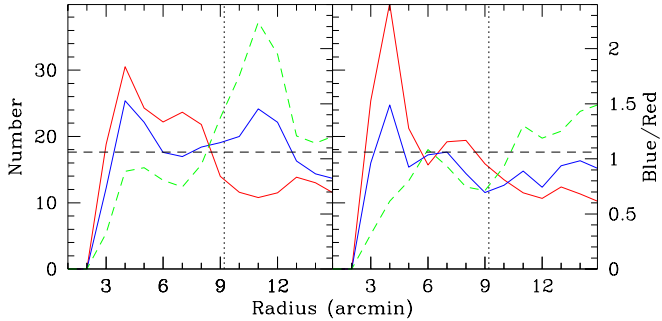}
   \caption{Radial trend of the galaxy population in the filaments, in bins of 1$^\prime$ width (left: S filament; right: NW filament. The galaxy number reads on the left margin, the blue-to-red ration on the right one). Along with the trend of the two populations (red and blue lines), the ratio of blue-to-red galaxies is shown (dashed green line) and compared with the mean value in the field (horizontal dashed line). Beyond the virial radius (vertical dotted line), a sudden increase in the ratio is seen, up to twice the field value.}
    \label{Fig.6}
    \end{figure}

Finally, we note that the relative fraction of blue galaxies increases
moving outwards along the regions containing the filaments (see Fig. 4),
possibly because the number of blue galaxies decreases less rapidly
than the number of red galaxies.
However, in the southern region a sudden increase of blue galaxies
contributes to the peak in the blue-to-red galaxy number ratio
at a cluster-centric distance of about $1.2~R_{\rm Vir}$.
This peak is significant with respect to the relative fraction
of blue galaxies in the field.

\section{Discussion and conclusions}

In conjunction with the seminal study of B\"ohringer et al.
(\cite{boehringer06}), our analysis strongly suggests
the presence of two large-scale filaments associated with the massive merging cluster
A\,2744 at $z = 0.3068$.
Similar evidence exists for other clusters at intermediate redshifts
(Kodama et al. \cite{kodama01}; Ebeling et al. \cite{ebeling04}).
On the other hand, the existence of a clumpy, large-scale filamentary
structure is at the basis of the current paradigm of cosmic formation
and evolution of clusters (e.g., Colberg et al. \cite{colberg99}).

We investigate star formation in galaxies selected in two regions
encompassing the large-scale filaments of A\,2744
and in an equivalent, wide region
where the density of galaxies at the (photometric) redshift of the cluster
is significantly lower (the field).
The overdensity in the filaments is driven by the population of galaxies
fainter than $\rm R^{\star}$ (as determined by Busarello et al.
(\cite{busarello02}) in the main body of the cluster),
whatever their B-R colour.
However, the blue-to-red galaxy number ratio increases
with increasing cluster-centric distance along the filaments
but peaks at about $1.2~R_{\rm Vir}$.
This result is consistent with and complementary to the established existence
of a radial dependence of azymuthal averages of cluster galaxy properties
like colours or emission-line strengths
(e.g. Balogh et al. \cite{balogh99}; Lewis et al. \cite{lewis02};
G\'omez et al. \cite{gomez03}; Finn et al. \cite{finn05}).
It is also consistent with the result that the relative fraction
of blue galaxies peaks between 1 and $2~R_{\rm Vir}$ in the nearby Shapley
and Pisces-Cetus superclusters (Haines et al. \cite{haines06};
Porter \& Raychaudhury \cite{porter07}).

A\,2744 is a merging cluster, where the mass-ratio
of the sub-components is about 3:1 (Boschin et al. \cite{boschin06}).
A qualitative comparison between the {\em Chandra} X-ray map (Kempner \& David \cite{kempner04}),
exhibiting a signature of a bow-shock,
and recent model renditions of shock heating in cluster mergers
(McCarthy et al. \cite{mccarthy07}) suggests that merging
started less than 2 Gyr before the observed epoch.
The primary shock could not travel to distances as large as the virial radius of the cluster.
Hence the galaxy population along the filaments of A\,2744
has not been affected by this large-scale merging event.

In their seminal spectroscopic study of the galaxy population
in the core region of this cluster, Couch et al. (\cite{couch98})
find that the majority of star-forming galaxies is generally made of systems
involved in major mergers but of modest luminosity,
even in this brightened phase.
These galaxies are interpreted as the progenitors of dwarf systems
once they fade.
Interestingly, we find that the filaments of A\,2744
host a population of luminous, large (i.e. brighter than $\rm R^{\star}$
and with sizes of 13--22 $h_{70}^{-1}$ kpc), blue galaxies
that hardly exists in the field.
Hence these galaxies can either retain or gain gas while they are falling
into the main body of the cluster.

Galaxy harassment (Moore et al. \cite{moore96}) is a suitable physical explanation for the enhanced star formation inferred for these sparse, luminous, blue galaxies found at distances close to the virial radius of A\,2744, where the density of the ICM is low.
As envisioned by Moore et al. (\cite{moore99}), high-surface brightness
disc galaxies and galaxies with luminous bulges do not experience
a significant removal of material nor a transformation in Hubble type
under the influence of high-speed, close encounters with substructure
(including bright galaxies) and strong tidal shocks from the global
cluster potential.
However, their discs will be heated and undergo instabilities
that can funnel gas (if any initially) into the central regions
(Lake et al. \cite{lake98}).
Galaxy harassment can generate morphological instabilities in galaxies already at the outskirt of a cluster (Moore et al. \cite{moore98}; Mastropietro et al. \cite{mastropietro05}).
In the core of the cluster, ram pressure can strip all the residual gas
from discs and drive a morphological transformation into S0's, as speculated
by Moore et al. (\cite{moore99}).
Alternatively, this can happen there once the shock associated
with the large-scale merging decelerates and discs can cross it
(see Roettiger et al. \cite{roettiger96}).
Therefore, the luminous, blue galaxies along the filaments of A\,2744
can be analogous to the progenitors of the S0--Sb galaxies
in the core of the cluster that completed their last major episode
of star formation 1--2 Gyr before the observed epoch
(Couch et al. \cite{couch98}).

\begin{acknowledgements}
This research has made use of the NASA/IPAC Extragalactic Database (NED)
which is operated by the Jet Propulsion Laboratory, California Institute
of Technology, under contract with the National Aeronautics
and Space Administration. D.P. acknowledges a stimulating discussion with
G. Lake. F.B. wants to thank A. Biviano for useful discussions and suggestions.
\end{acknowledgements}

\end{document}